\documentclass{aastex}
\usepackage{emulateapj5}

\shorttitle{Chandra Observation of Abell 262}
\shortauthors{Blanton, Sarazin, McNamara, \& Clarke}

\slugcomment{Astrophysical Journal, in press}

\begin{document}

\title{{\it Chandra} Observation of the Central Region of the Cooling Flow 
Cluster Abell 262:  A Radio Source that is a Shadow of its Former Self?}

\author{Elizabeth L. Blanton\altaffilmark{1,2},
Craig L. Sarazin\altaffilmark{1}, Brian R. McNamara\altaffilmark{3}
and T. E. Clarke\altaffilmark{1}}

\altaffiltext{1}{Department of Astronomy, University of Virginia,
P. O. Box 3818, Charlottesville, VA 22903-0818;
eblanton@virginia.edu, sarazin@virginia.edu, tclarke@virginia.edu}

\altaffiltext{2}{{\it Chandra} Fellow}

\altaffiltext{3}{Department of Physics \& Astronomy, Ohio University,
Clippinger Labs, Athens, OH 45701;
mcnamara@helios.phy.ohiou.edu}

\begin{abstract}
We present a {\it Chandra} observation of the cooling flow cluster Abell
262.  Spectral fits show that the intracluster medium (ICM) in A262 cools by
a factor of three from 2.7 keV to 0.9 keV at the cluster center.  A
mass deposition rate of $\dot{M} = 19^{+6}_{-5}~M_{\odot}$ yr$^{-1}$ is 
measured.  Complex structure is found in the very inner regions of the
cluster, including knots of emission and a clear deficit of
emission to the east of the cluster center.  The bright X-ray structures are
located in the same regions as optical line emission, indicating that
cooling to low temperatures has occurred in these regions.  The
X-ray deficit is spatially coincident with the eastern radio lobe associated 
with the active galactic nucleus hosted by the central cD galaxy.  The region 
surrounding the X-ray hole is cool, and shows no evidence that it has been 
strongly shocked.  This joins the ranks of other
cooling flow clusters with {\it Chandra}-detected bubbles blown by 
central radio sources.  This source is different than the other well-known
cases, in that the radio source is orders of magnitude less luminous and has
produced a much smaller bubble.  Comparing the energy output of the radio
source with the luminosity of the cooling gas shows that energy transferred to
the ICM from the radio source
is insufficient to offset the cooling flow unless the radio source is 
currently experiencing
a less powerful than average outburst, and was more powerful in the past.
\end{abstract}


\section{Introduction}

Recent X-ray observations from the {\it Chandra} and {\it XMM-Newton}
Observatories have shed much light on the physical state of the 
intracluster medium (ICM) in clusters of galaxies.
Gas is predicted to cool first in the dense centers of clusters of
galaxies since the cooling time varies as $t_{\rm cool} \propto T^{1/2}/n_e$,
where $T$ is the gas temperature and $n_e$ is the electron number density.
In order to maintain hydrostatic
equilibrium, gas then flows into the cluster center to make up for 
the
pressure lost from the cooling gas.  Evidence for these ``cooling flows''
(see Fabian et al.\ 1994 for a review)
was seen with observations with the {\it Einstein} (White 1992), {\it ROSAT} 
(Peres et al.\ 1998),
and {\it ASCA} (White 2000) observatories.
These observations predicted hundreds of solar 
masses of gas per year would flow into the centers of clusters and cool.
The main problem with this picture was that sufficiently large quantities
of cool gas were not inferred from observations taken at other wavelengths.
While increased star formation was observed in the centers of cD galaxies
in the centers of cooling flow clusters, as compared with non-cooling flows,
the rates measured did not match the cooling rates derived from the
X-ray observations (McNamara 1997).

Analysis of high-resolution spectroscopic data from {\it XMM-Newton} has
revealed that the majority of the cooling X-ray-emitting ICM in the centers
of cooling flows 
is cooling only over a limited temperature range, down to approximately
one-half or one-third of the temperature in the outer regions of the cluster (Peterson
et al.\ 2003).  Temperature profiles measured with {\it Chandra} data have
shown a similar phenomenon (McNamara et al.\ 2000; Johnstone et al.\ 2002;
Blanton et al.\ 2001, 2003).
In addition, the mass-deposition rates measured with both {\it XMM-Newton} 
and {\it Chandra} are lower than the rates found with the earlier
observatories.
Various scenarios have been proposed to explain the new results
including thermal conduction, inhomogeneous abundances, mixing, and heating of
the cooling ICM by a central radio source (see Fabian et al.\ 2001;
Peterson et al.\ 2001).

Remarkable high-resolution images from the {\it Chandra}
X-ray Observatory have revealed in great detail the profound effect that 
radio sources associated with the central bright elliptical galaxies have on 
the X-ray-emitting ICM.  The radio sources evacuate cavities (or ``blow
bubbles'') in the ICM, creating holes in the X-ray emission (e.g. Hydra
A, McNamara et al.\ 2000; Perseus, Fabian et al.\ 2000; A2052, Blanton et al.\
2001).  These bubbles rise buoyantly, transporting energy and magnetic fields 
into the ICM (as in Perseus, Fabian et al.\ 2000; A2597, McNamara et al.\ 
2001).  Theoretical models had predicted that the radio sources would
directly heat the ICM through strong shocks (Heinz, Reynolds, \& Begelman
1998).  However, in all observations of cooling flow clusters with central
radio sources to date, no evidence of strong-shock heating has been found.
In fact, the dense rims around the bubbles have been found to be cool, not 
hot, as would be predicted from a strong shock (Schmidt et al.\ 2002; Nulsen 
et al.\ 2002; Blanton et al.\ 2003).  More recent models (Reynolds, Heinz,
\& Begelman 2002) have found that weak shocks occurring some time in the
past are consistent with the observations (see also Soker, Blanton, \& 
Sarazin 2002).  
For example, recent observational evidence of weak shock
features have been found in Perseus (Fabian et al.\ 2003) and
M87/Virgo (Forman et al.\ 2004).
Comparisons of the total power emitted from the central radio source
with the luminosity of cooling gas have shown that the energy emitted
from a central radio source is sufficient to offset the cooling of the
ICM in the centers of clusters, at least in some cases (e.g. Hydra A, David
et al.\ 2001; A2052, Blanton et al.\ 2003).  The details of the transportation
of the radio source energy to the ICM are somewhat unclear thus far;
however, the scenario
of radio source heating remains a promising solution for the problem
of the missing cool gas in the cooling flow model.

In this paper, we present {\it Chandra} data of the
cooling flow cluster Abell 262.  We focus on the central region of the cluster
and the interaction between the radio source and the X-ray-emitting
intracluster medium.  Larger-scale properties of the cluster will be presented
in a future paper (Blanton et al., in preparation).  Abell 262 is a member 
of the Perseus
supercluster, along with the Perseus cluster (Abell 426) and Abell 347.
It is at a redshift of 0.0163 and is a richness class 0 cluster.  The 
central cD galaxy, NGC 708,
is host to a double-lobed radio source, B2 0149+35 (Parma et al.\ 1986;
Fanti et al.\ 1987).  A262 was previously observed in the X-ray with
{\it Einstein} (David et al.\ 1993), {\it ROSAT} (David, Jones, \& Forman
1996; Peres et al.\ 1998; Neill et al.\ 2001), 
{\it ASCA} (White 2000), and {\it XMM-Newton} (Peterson et al.\ 2003).
The {\it ASCA} data gave an average temperature of $kT = 2.3$ keV and
an average chemical abundance of 0.3 times the solar value.  The 
{\it Einstein}
data gave a bolometric luminosity of $L_{ X,{\rm bol}} = 4.4 \times 10^{43}$ erg
s$^{-1}$, scaled to our cosmology (see below).  Previous values
of the mass deposition rate are $\dot{M} = 13^{+14}_{-13}~M_{\odot}$ 
(spectral fit with {\it ASCA}; White 2000) and 
$\dot{M} = 14^{+2}_{-2}~M_{\odot}$ 
(morphological determination from {\it ROSAT}, Peres 1998), scaled to our 
cosmology.  Using {\it XMM-Newton} data, Peterson et al.\ (2003)
found $\dot{M} = 10\pm1~M_{\odot}$ in the range $kT$ to $kT/2$ where 
$kT = 2.1\pm0.2$ is the temperature of the ICM outside of the cooling region.
We assume $H_{\circ} = 70$ km s$^{-1}$ Mpc$^{-1}$, $\Omega_M = 0.3$,
and $\Omega_{\Lambda} = 0.7$ throughout.  At $z = 0.0163$, $D_L = 70.7$ Mpc
and $1\arcsec = 0.3318$ kpc.
Unless otherwise noted, all error bars are 90\% confidence regions.

\section{Observation and Data Reduction}

Abell 262 was observed with the {\it Chandra} X-ray Observatory on
2001 August 3 for a total of 28,743 seconds.  The cluster center was 
positioned one arcmin away from the nominal pointing of the ACIS-S3 detector
to avoid a node boundary on the chip.  The events were telemetered in very
faint (VF) mode, with an energy filter of 0.1-13 keV to avoid saturation
during telemetry.  The data were collected with frame times of 3.2
seconds and the CCD temperature was $-120$ C.  Only events with {\it ASCA} 
grades of 0, 2, 3, 4, and 6 were included.  In addition to the S3, we 
received data from the S1, S2, S4, and I3 CCDs.  We include data from 
the S3 only in our analysis.  Data from the S1 are used to check for 
background flares.

We used the data analysis package CIAO v2.2 for the data reductions.
Background files were taken from the blank sky observations of 
M. Markevitch\footnote{http://asc.harvard.edu/contrib/maxim/bg/}
included in the CIAO calibration database (CALDB).  Since VF-mode blank-sky 
observations with the
ACIS-S3 were not available at the time of our analysis, we processed the
data in Faint (F) mode, to be consistent with the background data.
Bad pixels, bad columns, and columns next to bad columns 
and node boundaries were excluded.  For the analysis, event PI values 
and photon energies were determined using the acisD2000-08-12gainN0003.fits 
gain file.  We searched for background flares using data from the S1, since
the cluster emission fills the S3.  We used the script
lc\_clean to clean
the data in the same manner as done by M. Markevitch for the blank sky
fields.  This script calculates the mean background rate after binning the
data into bins with a maximum length of 259.28 seconds (as for the blank sky 
data) and clips data that are more that 1.2 times different than the mean.  A 
total of 383 seconds of data were excluded using this technique, leaving a 
total exposure of 28,360 seconds. 

\section{X-ray Image} \label{sect:image}

The entire field of view of the ACIS-S3 CCD is shown in Figure 
\ref{fig:xrayall}.  The image is in the 0.3 -- 10.0 keV band.  It is
unsmoothed and has not been corrected for background or exposure.
The cluster shows substructure in the center.  The central bright region
is roughly elongated NE to SW, and there is an X-ray deficit to the east of
the center.  This elongation is approximately aligned with the position
angle of the central cD galaxy, as seen in the optical band (see Fig.\
\ref{fig:dss}).

\vskip3.25truein
\includegraphics{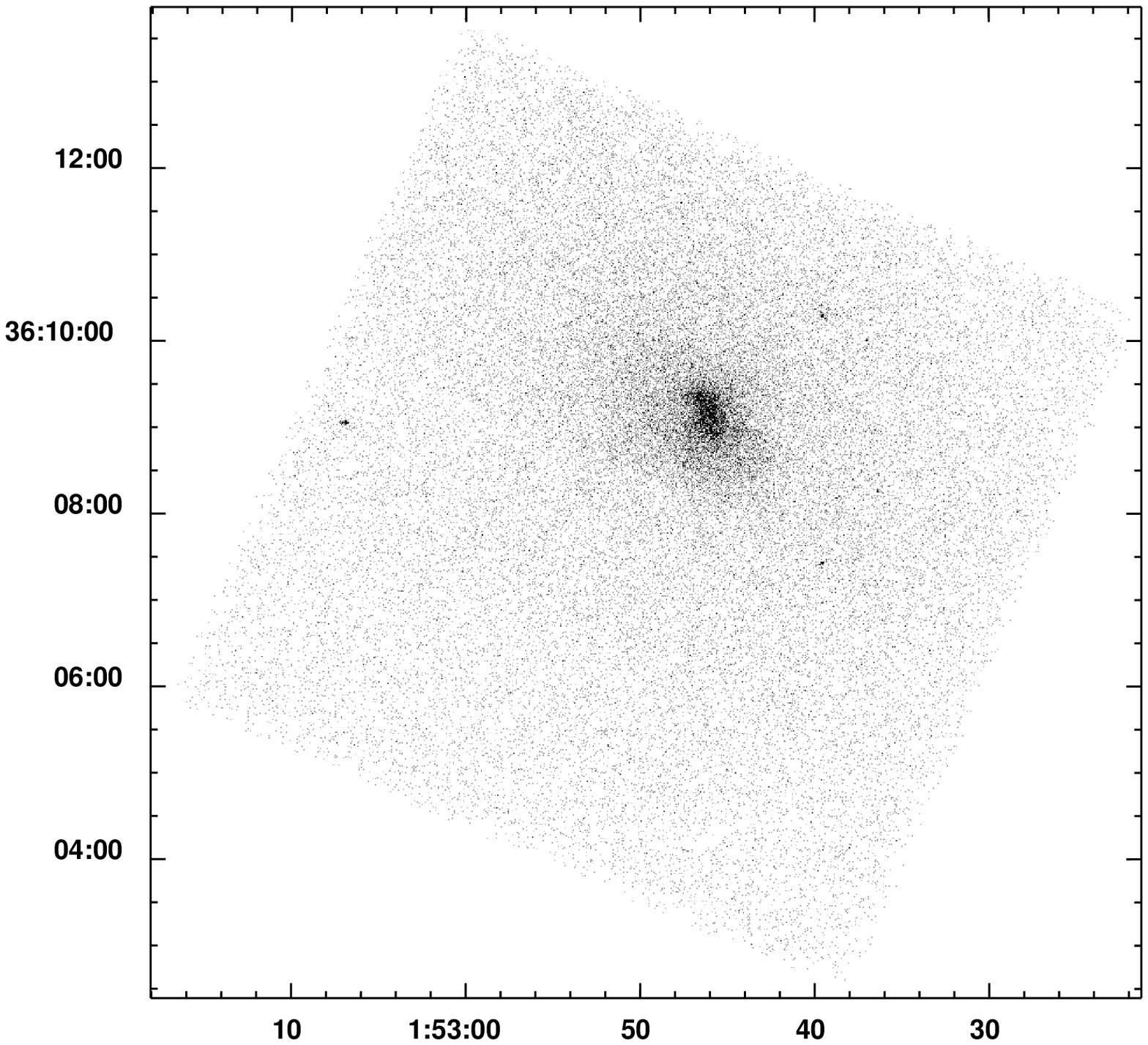}
\figcaption{Raw {\it Chandra} ACIS-S3 image of Abell 262.  The image is in the
0.3 -- 10.0 keV band and has not been corrected for background or exposure.
The central region is roughly elongated NE to SW, and there is a deficit in
the surface brightness to the east of the center.  The x and y axes are
RA (J2000) and Dec (J2000), respectively.\label{fig:xrayall}}
\vskip0.2truein

Individual sources were detected using the wavelet-detection algorithm
WAVDETECT in CIAO.  The source detection threshold was set at 
$10^{-6}$, implying that $\la 1$ false source (due to a statistical 
fluctuation) would be detected within the area of the S3 image.  Sources were
visually confirmed on the X-ray image.  All sources with a signal-to-noise
ratio (SNR) less than three were excluded from the final list, including 
low-level, very extended detections at the edges of the CCD.
The final list includes 17 sources, including four sources in the center of
the cluster.  Some of the sources in the center are likely regions of 
clumpy ICM rather than discrete X-ray sources.
Source positions were compared with optical positions from the USNO-A2.0
catalog (Monet et al.\ 1998).  One of the X-ray sources in the center is 
2\farcs68 arcsec away from
the USNO position of the central cD galaxy, NGC 708.
This same X-ray source is
$1\farcs81$ away from the radio core of B2 0149+35 as measured with
A-array VLA observations in Fanti et al.\ (1986).
Another X-ray source is $0\farcs4$ away from an object in the USNO catalog.
It is identified as NGC 703, a cluster galaxy to the NW of the central
galaxy, which is also a radio source, as seen in the NRAO VLA Sky Survey
(NVSS; Condon et al.\ 1998), with a flux density of 8.3 mJy.  There is one
additional match with an unidentified optical source in the western part of 
the field, with an offset of $1\farcs07$.
Since the offsets between the positions measured in the X-ray vs.\ those 
measured in the optical and radio are small and not systematic, we 
have not altered the 
astrometric solution for the {\it Chandra} data.
An optical image from the Second Generation Red Digitized Sky 
Survey\footnote{The Digitized Sky Surveys were produced at the Space 
Telescope Science Institute under U.S. Government grant NAG W-2166. The 
images of these surveys are based on photographic data obtained using the 
Oschin Schmidt Telescope on Palomar Mountain and the UK Schmidt Telescope. 
The plates were processed into the present compressed digital form with the 
permission of these institutions.} 
(DSS), trimmed to show the same
field-of-view as the ACIS-S3 image, is shown in Figure \ref{fig:dss}.
Detected X-ray sources are indicated with circles on the DSS image.  The
linear feature in the lower right is a defect on the DSS plate.

\vskip3.2truein
\includegraphics{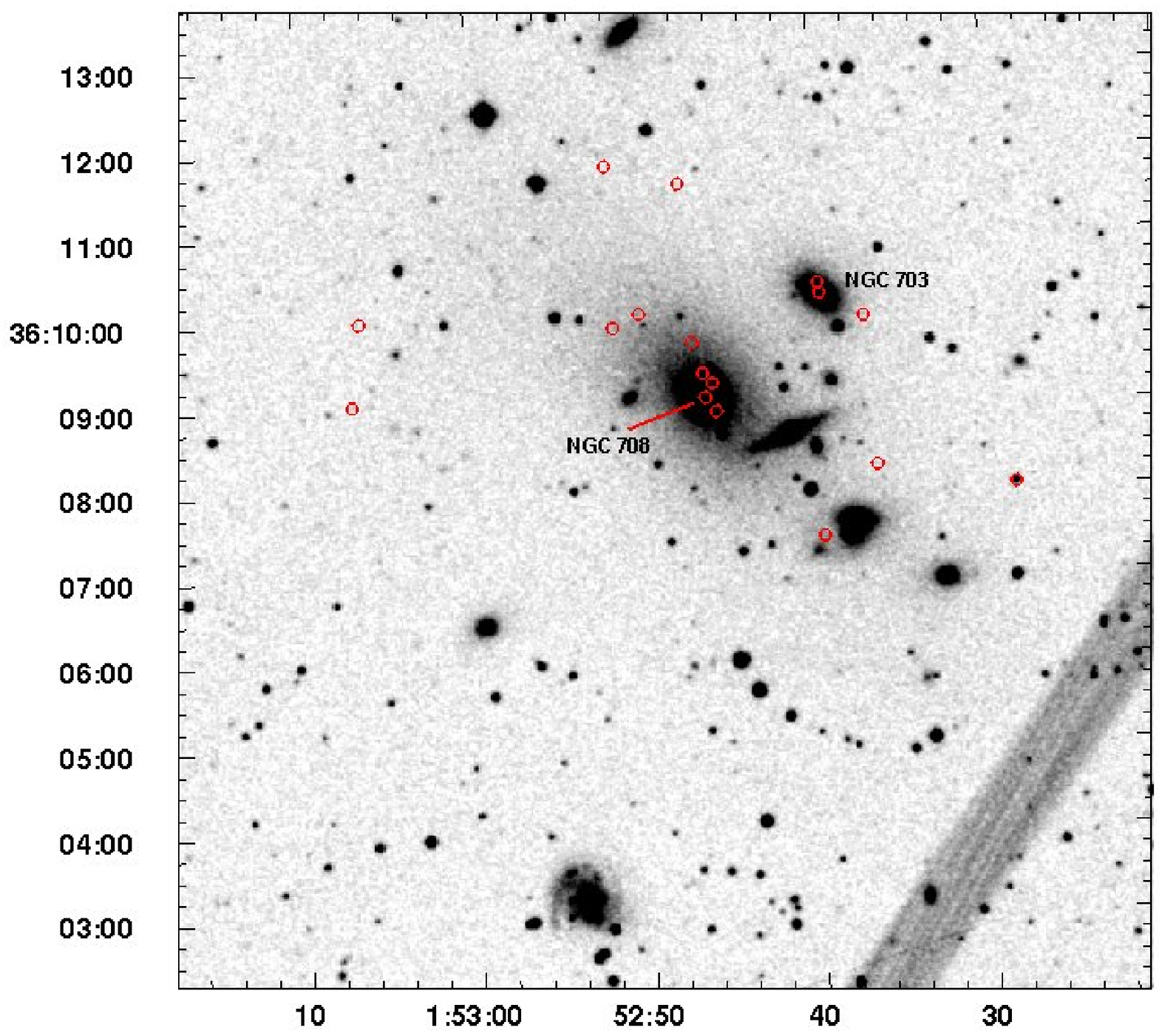}
\figcaption{Image from the Digitized Sky Survey (DSS), trimmed to match the 
field-of-view shown in Figure \ref{fig:xrayall}.  The positions of 
detected discrete X-ray sources are marked with circles.  The feature in 
the lower right corner is a defect on the DSS plate.\label{fig:dss}}
\vskip0.7truein

\section{Central Region of the Cluster}

An adaptively-smoothed image of the central $512 \times 512$ pixel
($252 \times 252$ arcsec; $84 \times 84$ kpc) region of the
cluster is shown in Figure \ref{fig:smoo}.  This image was created with the 
CIAO task CSMOOTH using a minimum signal-to-noise ratio of 3 per 
smoothing beam.  It was corrected for background and exposure, with the
background data taken from the blank-sky images in the CALDB, contributed
by M. Markevitch.

\vskip3.2truein
\includegraphics{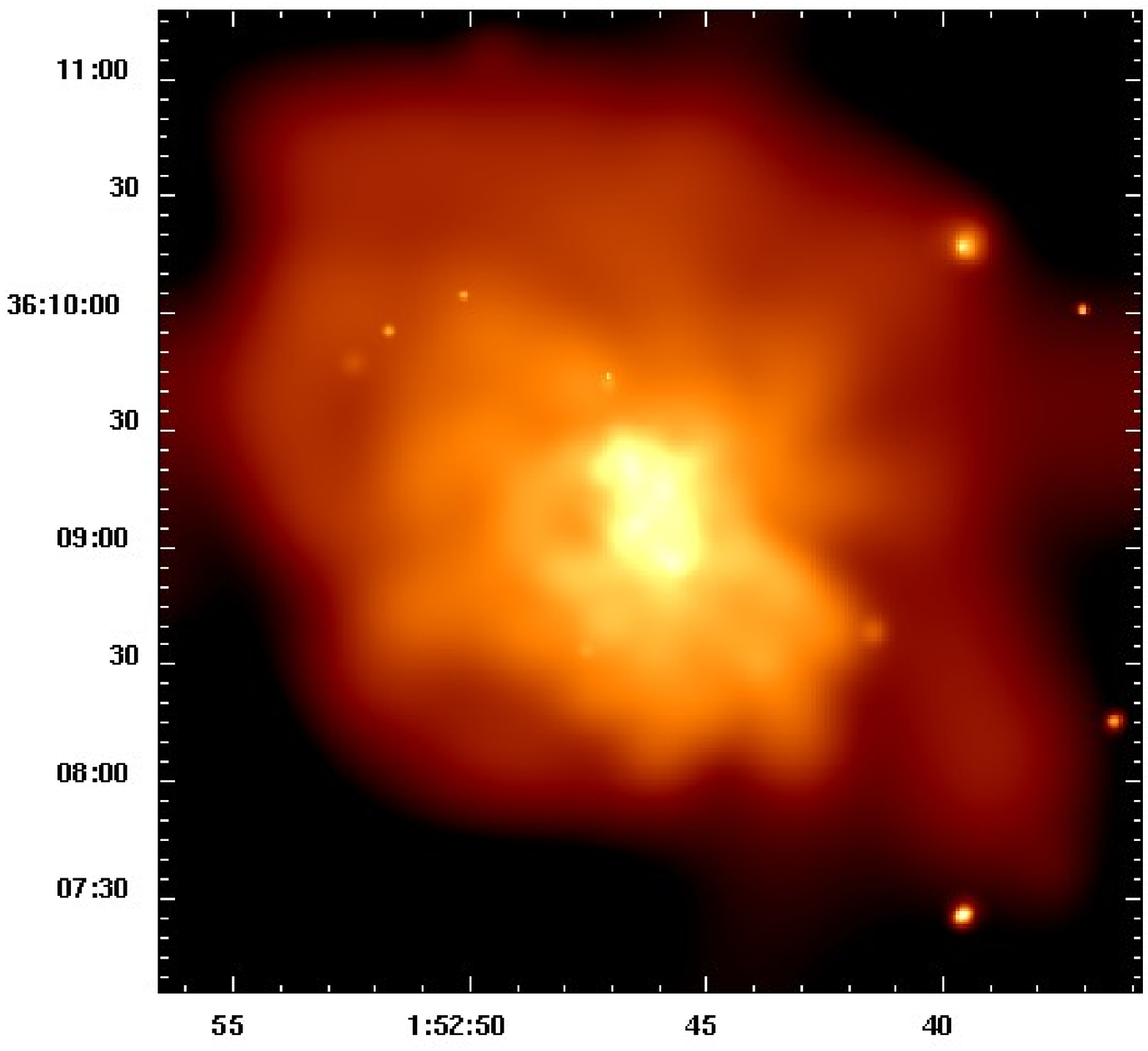}
\figcaption{Adaptively-smoothed image of the central $252 \times 252$ arcsec
($84 \times 84$ kpc) region of Abell 262.  The image has been corrected for 
background and exposure.  Several structures are visible in the image,
including a clear ``hole'' or ``bubble'' in the emission approximately
$10\arcsec$ east of the cluster center.\label{fig:smoo}}
\vskip0.2truein

The image shows a variety of structures, including individual point sources,
bright knots and filaments of emission, and regions of surface brightness
deficit, including a clear ``hole'' or ``bubble'' in the emission just east
of the center of the cluster.  The central
approximately $30\arcsec$ region contains four bright knots of emission,
including a source coincident with the radio core, which may be due to the
AGN.  To examine this possibility, we have determined the hardness
ratios for the ``AGN'' component and the sum of the three other knots.
We define the hardness ratio as (H-S)/(H+S) where H is the number of
counts in the $1 - 10$ keV range, and S is the number of counts in the 
$0.3 - 1$ keV range.  Using the blank-sky background data to subtract the
background, we find
a hardness ratio of $0.14\pm0.08$ for the possible AGN and $-0.13\pm0.05$
for the sum of other three knots.  Therefore, the hardness ratio for the
former is significantly harder than the latter, and probably is due to 
an additional non-thermal, hard emission component from the AGN.  These ratios
include all emission along the line of sight towards the AGN and the knots,
since the blank sky files were used for the background.  If we instead take
the background locally in annuli surrounding the extraction apertures for
each region of interest, we find hardness ratios of $1.0^{+0.0}_{-2.0}$ for
the AGN and $-0.40^{+0.53}_{-0.35}$ for the sum of the other knots.  The
errors are much larger in this case, since the net counts are greatly reduced
when subtracting the background locally, and in fact, when doing this, the
net soft counts for the AGN are only $0.04\pm9.7$.  Finally, if we examine
the hard ($1 - 10$ keV) and soft ($0.3 - 1$) keV images, side by side, it
is evident that the AGN component includes hard emission that the other
knots do not (see Figure \ref{fig:hardness}).

\vskip2.1truein
\includegraphics{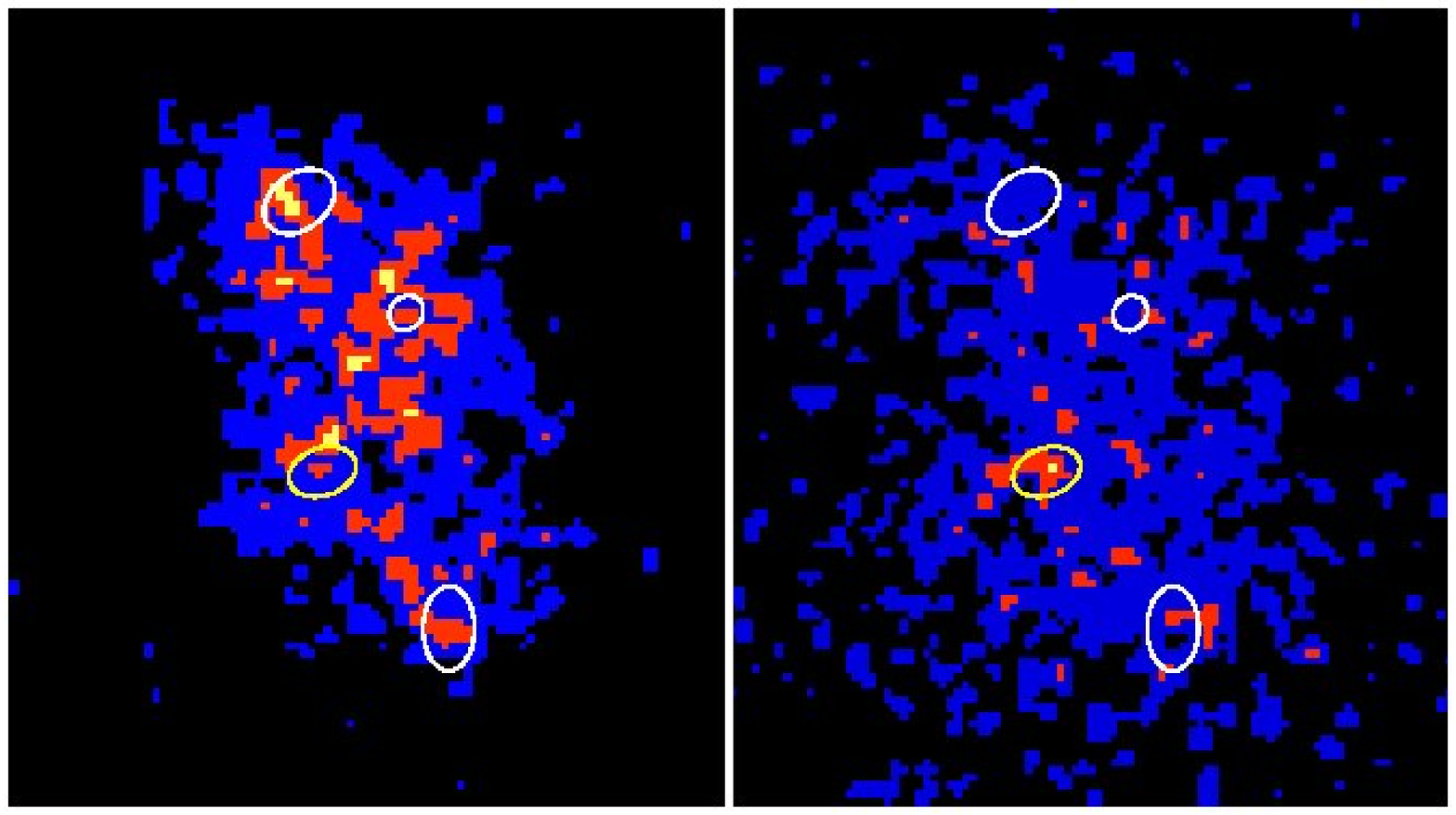}
\figcaption{Slightly (1 pixel) Gaussian-smoothed images in the soft ($0.3 - 1.0$
keV; left panel) and hard ($1.0 - 10.0$ keV; right panel) bands of the central
$45\arcsec \times 50\arcsec$ region of Abell 262.  The positions of the AGN
and the other bright knots are marked with yellow and white elliptical 
regions, respectively, as found with WAVDETECT.  The AGN is significantly 
harder than the other knot regions.\label{fig:hardness}}
\vskip0.2truein

\vskip3.1truein
\includegraphics{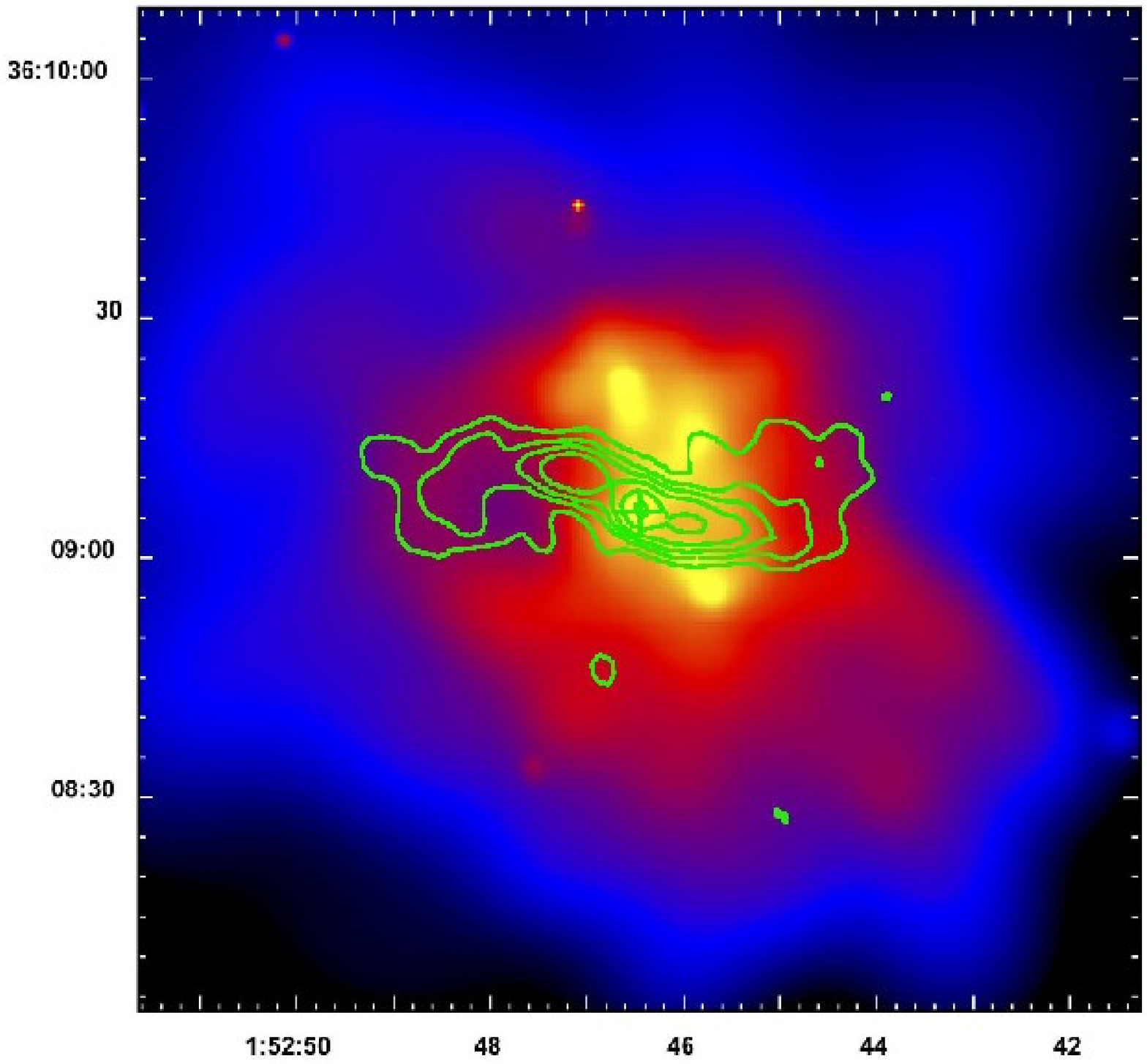}
\figcaption{Adaptively-smoothed {\it Chandra} image of the central $125 \times
125$ arcsec region of Abell 262 with radio contours at 1.4 GHz (Parma et al.\
1986) overlaid.  The radio emission is generally anti-coincident with the
X-ray emission; this is most clearly seen to the east of the cluster center, 
where the radio lobe has evacuated a cavity in the X-ray-emitting ICM.
\label{fig:radio}}
\vskip0.2truein

\subsection{Radio Source / ICM Interaction}\label{sect:rad}

The 1.4 GHz radio contours of B2 0149+35 (Parma et al.\ 1986) are overlaid
onto the adaptively-smoothed {\it Chandra} image of the central ($125 \times
125$ arcsec) region of Abell 262 in Figure \ref{fig:radio}.  The radio data
are from the VLA in B configuration and have a resolution of $\sim 3\farcs5$.
The total 1.4 GHz flux from the radio source is 78 mJy.  This gives a 
power at 1.4 GHz of $P_{1.4} = 4.7 \times 10^{22}$ W Hz$^{-1}$, classifying
the source as a fairly weak, double-lobed FR I (Fanaroff \& Riley 1974).  The 
largest-angular-size
of the radio source is $58\arcsec = 19.2$ kpc.
The radio source is much less luminous and physically 
smaller than other similar objects that have had their environments 
well-studied with {\it Chandra} data, such as Hydra A (McNamara et al.\ 2000), 
3C 84 in the Perseus cluster (Fabian et al.\ 2000), and 3C 317 in Abell 
2052 (Blanton et al.\ 2001, 2003).  

In Figure \ref{fig:radio}, an anti-correlation between the radio and X-ray 
emission is apparent.  This is most clearly
seen to the east of the cluster center where the radio lobe has ``blown a
bubble'' in the ICM.  The radio lobe fills a region of low-surface brightness
in the X-ray image and is surrounded by an X-ray-bright shell of emission.
This bubble has an inner diameter of $15\farcs5 = 5.2$ kpc.
The AGN is visible as a point source in the X-ray that is coincident with
the radio core.

\subsection{Correlation with [\ion{N}{2}] Emission}\label{sect:NII}

Contours of [\ion{N}{2}] optical line emission at $6584~\rm\AA$  are 
superposed onto
the adaptively-smoothed {\it Chandra} image of Abell 262 in Figure
\ref{fig:NII}.  The optical data are from Plana et al.\ (1998).
Overall, the [\ion{N}{2}] emission is positively correlated with
the X-ray emission, and the [\ion{N}{2}] emission has extensions that
correspond with the X-ray-bright knots of emission in the center of the
cluster.
The [\ion{N}{2}]
emission represents gas at a temperature of approximately $10^{4}$ K,
and may indicate that at least some of the gas is cooling to low temperatures
in these regions.  In addition, this argues that the emission from the
X-ray knots is thermal in nature.
Similar correspondences between X-ray and optical line emission have
been seen in the Abell 2052 (Blanton et al.\ 2001) and Abell 1795 (Fabian
et al.\ 2001b) clusters.

\vskip3.1truein
\includegraphics{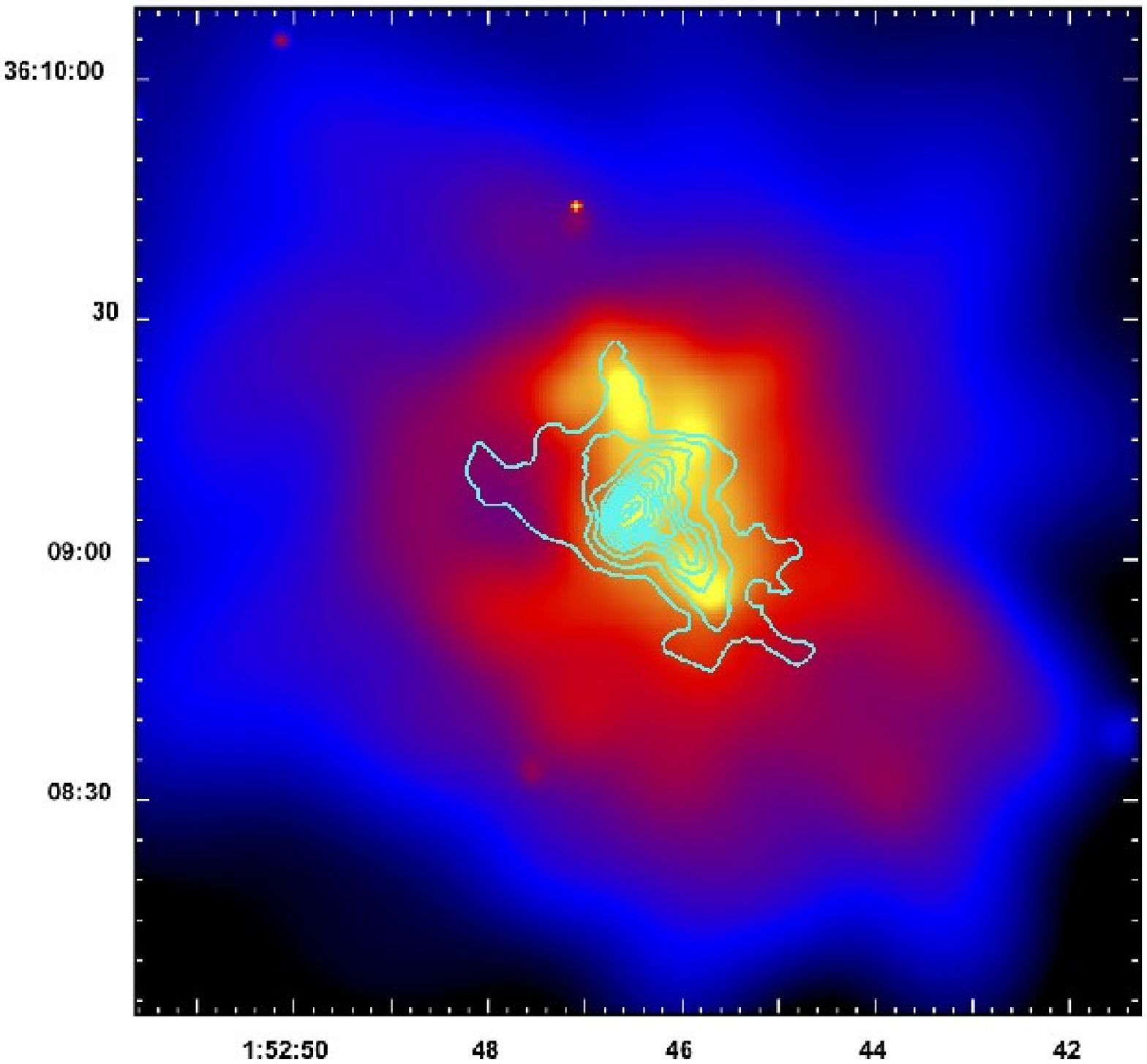}
\figcaption{Contours of [\ion{N}{2}] optical line emission superimposed onto the
adaptively-smoothed central region of the {\it Chandra} image of Abell 262.
In general, the [\ion{N}{2}] emission is positively correlated with the X-ray
emission, and particularly with the X-ray-bright knots, suggesting that
cooling to low temperatures is occurring in these regions.\label{fig:NII}}
\vskip0.2truein

\subsection{Temperature Map}

A temperature map covering the same region shown in Figures \ref{fig:radio}
and \ref{fig:NII} is displayed in Figure \ref{fig:tmap}.  This map was
obtained by fitting individual spectra with ISIS in the 0.7 -- 8.0 keV band, 
with the requirement that
each spectrum contain a minimum of 1050 counts, including background.
Ancillary response function (arf) files were corrected for the
low-energy quantum efficiency (QE) degradation in the ACIS-S3 using 
CORRARF.
Point sources were removed before creating the map.  The three thermal
``knots'' of emission that were detected with WAVDETECT were kept in the
image, since they most likely represent dense regions of gas, rather than
point sources.
When constructing the map,
each spectrum was binned to a minimum of 20 cts/bin before fitting with a
model including Galactic absorption fixed to 
$5.46 \times 10^{20}$ cm$^{-2}$ (Dickey \& Lockman 1990), and a thermal
plasma model (WABS*MEKAL).  The map is an array of $25 \times 25$ boxes
covering a total of $256 \times 256$ pixels.  The size of each box is 
$5\arcsec \times 5\arcsec$.  Each box does not represent the total size of 
the region 
covered by an individual spectral fit, but rather, the center of the region
from which a spectrum was extracted and fitted, with the requirement that
the region contain at least 1050 counts.

\vskip3.1truein
\includegraphics{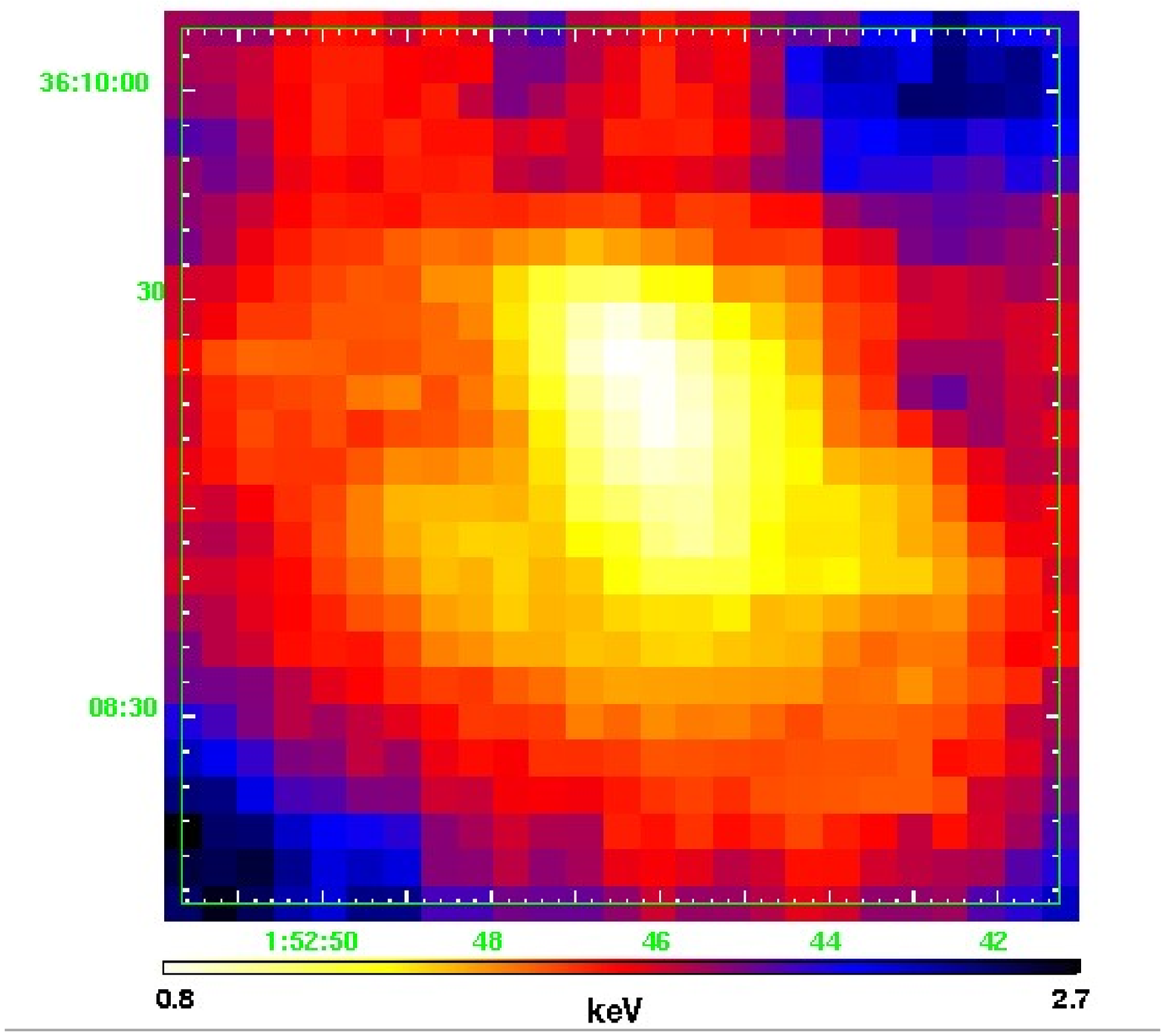}
\figcaption{Temperature map of the central $125 \times 125$ arcsec region of
Abell 262.  The gas is coolest in the center, with a temperature of 0.8 keV,
and rises to 2.7 keV farther out; the color scale shown is linear.  The 
gas surrounding the radio bubble to the east of the center is cool, and 
shows no sign of being shocked.
\label{fig:tmap}}
\vskip0.2truein

The map shows a range of temperatures from 0.8 to 2.7 keV, with the coolest
gas found in the center of the cluster, consistent with the temperature
distribution predicted from cooling flow models.
Errors on the temperature values are typically $10 - 15\%$.
Cool gas is found in the rim of the bubble 
evacuated from the radio source to the east of the cluster center.
This is inconsistent with the temperature distribution expected if the
radio lobe was strongly shocking the ICM.  Therefore, there is no evidence
that the radio lobes are currently providing heat to the ICM through
strong shocks, although energy may still be transported through slower 
expansion of the lobes (see Reynolds, Heinz, \& Begelman 2002).  Similar
distributions in the gas temperature have been seen in 
Perseus (Schmidt et al.\ 2002), Hydra A (Nulsen et al.\ 2002), and 
Abell 2052 (Blanton et al.\ 2003).

\section{Total Spectrum}

A spectrum was extracted from the largest circular region that would fit on
the ACIS-S3, centered on the core of the radio source.  The region has a
radius of $182\arcsec$ (60.4 kpc).  
Point sources detected with WAVDETECT, as
described in \S\ref{sect:image}, were excluded.  The source coincident with
the radio core was excluded, but the three other bright knots of emission
in the center were included, since they are thought to be thermal in nature.
The spectrum was extracted, and weighted
response functions for the region were calculated using the CIAO ACISSPEC
script.  Since this script only became available with CIAO 2.3, we use this
version of CIAO for the spectral analysis.  The gain file we used remained
unchanged from CIAO 2.2 to CIAO 2.3.
The spectral data were grouped to have a minimum of 20 cts/bin.  The CIAO
script ACISABS was applied to correct for the low-energy quantum efficiency 
(QE) degradation in the ACIS-S3.
The blank-sky observations included in the CALDB and contributed by M.
Markevitch were used to extract a background spectrum in the same manner and
from the same region described above.

A number of fits were made to the spectrum using XSPEC v11.2.0bg.  We fitted
single-temperature (1-MEKAL) thermal plasma models, two-temperature 
(2-MEKAL) models, and
cooling flow models (MKCFLOW+MEKAL).  In each case, the absorption was fixed 
to the Galactic value of $5.46 \times 10^{20}$ cm$^{-2}$ (Dickey \& Lockman
1990), and then allowed to vary.  We fitted the spectrum in the 0.7 -- 8.0
keV range, excluding the 1.8 -- 2.1 keV region that is adversely affected
by a feature produced by the iridium coating of {\it Chandra}'s mirrors.  
A total of 77,972 counts were detected in the spectrum in the energy range we
used for our fits.
A systematic error of 2\% was added to all of the fits to include the
effects of uncertainties in the calibration.
The results of the spectral fits are summarized in Table 1.

\vskip2.5truein
\includegraphics{f8.eps}
\figcaption{Total spectrum for Abell 262, extracted from a circular region
with a radius of $182\arcsec$ (60.4 kpc) centered on the radio core.
The model shown fitted to the spectrum is a cooling flow model with Galactic
absorption (MKCFLOW+MEKAL-1), with parameters as listed in Table
1.  The 1.8 -- 2.1 keV region has been excluded because it
is affected by an iridium feature resulting from {\it Chandra}'s mirror 
coating.\label{fig:totspec}}
\vskip0.2truein

For the single-temperature fit (1-MEKAL), the average temperature of the
cluster is $kT = 1.77^{+0.025}_{-0.080}$ keV, and the average abundance is
$0.40^{+0.016}_{-0.022}$ times the solar value.  However, this fit clearly 
does not adequately describe
the data.  The reduced $\chi^2$ value is 2.60.  The fit is improved when
the absorption is allowed to vary (1-MEKAL-abs).  This results in an 
absorbing column which is lower than the Galactic value, which we consider
unlikely. 
The low fitted absorption may be due to uncertainties in the correction
for the low-energy QE degradation or to additional cooler emission
components.
The fit is significantly improved with the addition of a second thermal
component (2-MEKAL).  In this case, the best-fitting low and high 
temperatures are $kT_{\rm low} = 1.11^{+0.024}_{-0.035}$ keV, and 
$kT_{\rm high} = 2.36^{+0.074}_{-0.067}$ keV, respectively.
The average abundance is 
$0.67^{+0.055}_{-0.059}$ times the solar value.  The reduced $\chi^2$ value 
drops to 1.62 for this fit. The two-temperature fit is
slightly improved when the absorption is allowed to vary (2-MEKAL-abs).
Here, again, the best fitting value for the absorption is below the Galactic
value, but not as much so as for the single-temperature fit.

We fitted the spectrum with a series of cooling flow models, that also 
include an additional MEKAL component to account for the projection of any
outer non-cooling gas in the cluster onto the cooling region.
In each case,
the higher temperature ($kT_{\rm high}$) in the cooling flow model (MKCFLOW)
was set to be equal to the temperature of the MEKAL component,
under the assumption that the MEKAL component represents ambient cluster gas,
and that the cooling flow gas is cooled ambient gas.
Similarly,
the abundance values for the two models were constrained to be
equal.  The fits were performed with the absorption both fixed to the Galactic
value and allowed to vary, as above.  In addition, the low temperature 
component ($kT_{\rm low}$) was either allowed to vary, or fixed to a very
low temperature (0.08 keV) to constrain the amount of gas cooling down to
very low temperatures.  The fits were better when the low temperature 
component was allowed to vary freely.  These fits indicate that the gas
is cooling over only a limited temperature range, from approximately 2.7 to 
0.9 keV, similar to the range seen in the temperature map.  For the fixed 
absorption case (MKCFLOW+MEKAL-1), the best-fitting
lower and upper temperatures are $0.87^{+0.096}_{-0.047}$ keV and
$2.65^{+0.17}_{-0.18}$ keV, respectively, and the mass-deposition rate over
this limited temperature range is $\dot{M} = 18.8^{+5.7}_{-4.5}~M_{\odot}$ 
yr$^{-1}$.  The abundance for this fit is
$0.66^{+0.064}_{-0.063}$ times the solar value, and the reduced $\chi^2$ 
value (1.61) is only slightly better than the 2-MEKAL fit.  The 
free-absorption case (MKCFLOW+MEKAL-abs1) yields similar results, and an
absorption value consistent with the Galactic value within the errors.
When the low-temperature component is fixed to $kT = 0.08$ keV and
the absorption is fixed to Galactic (MKCFLOW+MEKAL-2),
the mass-deposition rate
drops to $\dot{M} = 5.0^{+0.5}_{-0.6}~M_{\odot}$ yr$^{-1}$.
This suggests that
approximately $19~M_{\odot}$ yr$^{-1}$ are cooling over the limited
temperature range of 2.7 to 0.9 keV, with only approximately $5~M_{\odot}$ 
yr$^{-1}$ cooling to very low temperatures (0.08 keV).
If the absorption is allowed to vary with the low temperature fixed to
0.08 keV (MKCFLOW+MEKAL-abs2), the mass-deposition
rate to very low temperatures increases to 
$\dot{M} = 11.7^{+1.8}_{-1.4}~M_{\odot}$ yr$^{-1}$.  However, this requires a
very high absorption value ($N_H = 14.1^{+1.79}_{-1.45} \times 10^{20}$ 
cm$^{-2}$) to absorb
out the expected emission predicted from the cool gas at low energies, and
the fit is significantly worse than the cooling flow models where the 
low-temperature component is allowed to vary.
The best fit model is MKCFLOW+MEKAL-1, where the absorption is
fixed to the Galactic value, and the gas cools over only a limited temperature
range from 2.7 to 0.9 keV.  A plot of this model fitted to the spectrum is 
shown in Figure \ref{fig:totspec}.  We note that this fit is only slightly 
better than the 2-MEKAL model.

\section{Radio Bubbles}

As described in Section \ref{sect:rad}, there is a depression in the
X-ray surface brightness to the east of the AGN.  This depression is
coincident with the eastern radio lobe.  The surface brightness in an
annular region surrounding the hole is $25\%$ higher, and brighter than
the hole at the $4.3\sigma$ level.
Visually, there seems to be some
indication for an additional hole approximately $27\arcsec$ east of this
one, which may represent a ``ghost cavity'' as seen in the Perseus cluster
(Fabian et al.\ 2000) and Abell 2597 (McNamara et al.\ 2001).  
These may be cavities evacuated from a previous outburst of the radio
source. However, this feature is not statistically significant.

We have extracted a spectrum from the outer (eastern) half of the 
bright shell surrounding the hole that corresponds with the eastern 
radio lobe.  The inner and outer radii of this region are 6\farcs4 (2.1 
kpc) and 17\farcs0 (5.6 kpc), respectively, centered on the hole.  
The opening angle used for the pie-annulus was 180 degrees.  
The background spectrum was extracted from a pie-annulus region centered on
the AGN and located 
just exterior to the bright shell in an attempt to subtract off emission 
along the line-of-sight to the shell from the cluster at radii larger than
the bright shell.
The spectral data were binned to a minimum of 20 cts/bin and the spectrum
was fitted with a model including absorption fixed to the
Galactic value and a MEKAL component with the abundance fixed to 0.6 times
solar.  The fit gave $\chi^{2}$/d.o.f. = 35.96/40 and a temperature of
$kT = 1.20^{+0.17}_{-0.19}$ keV.
Since the normalization of the model is proportional to $n_e^2$, we can
determine the density and pressure in this region.  We assume that the 
emission for the outer shell region is part of a spherical shell of emission.
We find an electron density of $n_e = 0.021$ cm$^{-3}$ and a pressure
$P = 8 \times 10^{-11}$ dyn cm$^{-2}$.  
Note that if we take our background spectrum from the blank-sky files in
the same region that we extract the shell spectrum, the ``projected'' pressure
is higher, $P = 1.8 \times 10^{-10}$ dyn cm$^{-2}$.  
Following a similar procedure as above, but extracting the spectrum from a 
full, circular shell surrounding the radio bubble, with inner and outer radii 
of 7\farcs8 (2.6 kpc) and 13\farcs0 (4.3 kpc), respectively, and 
extracting the background locally from a region just east of this shell, we 
find a pressure $P = 1.2 \times 10^{-10}$ dyn cm$^{-2}$.  We therefore adopt
a pressure value of $P = 1 \times 10^{-10}$ dyn cm$^{-2}$.
This is a factor of five higher than the pressure determined from the radio 
observations for the eastern radio lobe of 
$P_{\rm eq} = 1.9 \times 10^{-11}$ dyn cm$^{-2}$ (Heckman et al.\ 1989, converted
to our cosmology), 
assuming equipartition of energy.  This suggests that there is an additional 
pressure component in the radio bubble providing support to the X-ray shell.  
One possibility for this component is very hot, diffuse thermal gas.  A 
similar pressure discrepancy has been found in other similar objects such as
Hydra A (McNamara et al.\ 2000) and Abell 2052 (Blanton et al.\ 2001).
The isobaric cooling time in this portion of the bright shell is 
$4.0~(2.8) \times 10^{8}$ yr, for the partial (full) shell case, much shorter 
than the probable age of the cluster.  
This is consistent with the correlation of cool ($\sim10^4$ K) gas 
(see \S\ref{sect:NII}) with the X-ray-bright regions in the cluster center.
The synchrotron age of the radio source, calculated from the 1.4 GHz data
presented in Parma et al.\ (1986), is $3.5 \times 10^7$ yr, much shorter than
the cooling time.  Therefore, the gas in the shell did most of its cooling 
to its current temperature while it was closer to the center of the cluster, 
and was then pushed outward by the radio source.

\subsection{Energy Injection into the ICM}

One of the most promising solutions to the ``cooling flow problem''
(sufficient quantities of gas are not found at
sufficiently low temperatures, as based on expectations from X-ray data)
is that energy input from an AGN hosted by a central cluster galaxy can
offset the cooling.
Even without knowing the details of the mechanism by which this energy
is transferred to the ICM, we can test whether the energy output from the 
radio-emitting AGN is adequate to balance the cooling losses.  This has
been found to be the case in other objects such as
Hydra A (David et al.\ 2001) and Abell 2052 (Blanton et al.\ 2003).

The luminosity of isobaric cooling gas is given by
\begin{equation}
L_{\rm{cool}} = \frac{5}{2}\frac{kT}{\mu m_{p}} \dot{M} ,
\end{equation}
where $kT$ is the temperature of the ICM outside of the cooling region,
$\dot{M}$ is the cooling rate, and $\mu$ is the mean mass per particle in
units of the proton mass.
Using $kT = 2.65$ keV, and the mass-deposition rate of 
$18.8~M_{\odot}$ yr$^{-1}$, we find
$L_{\rm{cool}} = 1.3 \times 10^{43}$ erg s$^{-1}$.
We used the mass-deposition rate found in the spectral fit where the low
temperature was allowed to vary.  This represents gas cooling from 2.65 to
0.66 keV, and for the comparison with the radio source energy, we assume
that all of this mass needs to be heated.  Using a different method,
subtracting the spectroscopically-determined cooling luminosity (with the 
temperature fixed to a low value) from the morphological measure of the 
cooling luminosity, B\^irzan et al.\ (2004) find a very similar value 
($L_{\rm{cool}} = 1.2 \times 10^{43}$ erg s$^{-1}$) for A262.
We compare our value of the cooling luminosity with the total energy output 
from the radio source.
Following Churazov et al.\ (2002), the energy required to inflate the
bubble seen in the X-ray is the sum of the internal energy of the bubble
and the work done in creating the cavity and compressing the shell
\begin{equation}
E_{\rm rad} = \frac{1}{(\gamma - 1)}PV + PdV = \frac{\gamma}{(\gamma - 1)}PV
\, ,
\end{equation}
where $V$ is the volume of the cavity, and $\gamma$ is the mean adiabatic
index of the fluid in the cavity.
Using 7\farcs8 (2.6 kpc) as the radius of the hole to the east of the
cluster center, and the pressure derived above of $P = 1 \times 10^{-10}$ 
dyn cm$^{-2}$,
we find $E_{\rm rad} = 5.3 \times 10^{56}$ erg for nonrelativistic gas
($\gamma=5/3$, $E_{\rm rad}=5/2~PV$) and
$E_{\rm rad} = 8.5 \times 10^{56}$ erg for relativistic gas
($\gamma=4/3$, $E_{rad}=4PV$).
We assume that
the western radio lobe is supplying the same amount of energy as the 
eastern, giving a total energy for both lobes of 
$E_{\rm rad} = 1.1 \times 10^{57}$ erg ($E_{\rm rad} = 1.7 \times 10^{57}$ 
erg) for $\gamma=5/3$ ($\gamma=4/3$).
As evidenced by the presence of ghost cavities (holes in the X-ray emission
that have risen buoyantly outward from the cluster centers into the ICM) in 
other cooling flow clusters such as Perseus (Fabian et al.\ 2000) and 
Abell 2597 (McNamara et al.\ 2002), radio sources are episodic with a
repetition rate of $\approx 10^{8}$ yr (derived from the buoyancy rise time of
the cavities).
Assuming the same repetition rate of radio outbursts in the core of Abell
262, we find that the total power output from
the radio source is $3.4 \times 10^{41}$ erg/s ($5.4 \times 10^{41}$ erg/s)
for $\gamma=5/3$ ($\gamma=4/3$).
This falls more than an order of magnitude short of the necessary 
average energy output rate required
to offset the luminosity of cooling gas in Abell 262.
This result is different from that for other well-studied cases such as
Hydra A and Abell 2052, where $E_{\rm rad}$ is sufficient to balance
cooling.
Although the X-ray gas pressures in the centers of all three clusters are
similar,
the bubble volumes are much greater in Abell 2052 and Hydra A (bubble 
diameters $\approx 20$ kpc) than in Abell 262 (bubble diameter $\approx 5$
kpc).
In addition, the observed radio powers in the other objects are
much higher (by orders of magnitude) than in Abell 262.
It may be that some other mechanism (e.g., thermal conduction) balances
radiative cooling in Abell 262 rather than heating by the radio source.
Alternatively,
it is possible that the radio source in Abell 262 is experiencing
a less powerful than average outburst, and that previous
outbursts were more powerful allowing the radio source to balance
the cooling losses on average.
Another possibility is that the repetition
rate is higher for this radio source than others -- a repetition rate of 
$6 \times 10^{6}$ yr ($\gamma = 5/3$) or $1 \times 10^{7}$ yr ($\gamma = 4/3$)
rather than the assumed $10^{8}$ yr would be required.  
The problem with a rapid repetition rate is that, if the axis of the jets
remains fixed, this might simply add to the observed energy content of the
present radio bubble.
The required repetition time is shorter than the synchrotron
lifetime of the radio source of $3.5 \times 10^7$ yr, so repeated
injections into the same bubble would just add to the observed radio
emission.
If the buoyancy time of the bubbles were small enough, they might move
away between outbursts.
Taking the sound speed (500 km s$^{-1}$) as the upper limit of the
buoyancy velocity, the time for a bubble to move one bubble diameter 
(5 kpc) is $t_{\rm{buoy}} > 1 \times 10^7$ yr.
A radio source repetition rate that would repeat on a shorter 
timescale than this (as would be necessary in the above scenario where
a more rapid repetition rate supplies the required energy to offset the 
cooling) would result in more energy being injected into the 
current bubble which would be detectable
in the pressure and volume of the X-ray shell.  Therefore, it seems unlikely 
that a shorter repetition rate
would account for the necessary energy required to offset the cooling flow.
The more likely possibility is that previous outbursts of the radio source
were more powerful, if the radio source is to supply the required energy.

\section{Conclusions}

We have presented an analysis of the {\it Chandra} observation of the
central region of the cooling flow cluster Abell 262.  Complex structure is
found, including bright knots of emission and a hole
in the X-ray emission that is spatially-coincident with the eastern radio
lobe associated with the radio source hosted by the central cD galaxy.
One of the bright knots of emission is coincident with the core of the radio
source, and a hardness ratio analysis indicates that this source includes
an emission component from the AGN.
The knotty structure seen in the center is located in the same region as
optical [\ion{N}{2}] line emission, suggesting that gas is cooling down to
approximately $10^4$ K in these regions.  The temperature map shows that 
these same regions are among the coolest areas in the cluster, with 
X-ray-derived temperatures of approximately 0.8 keV.  The map also shows 
that the 
compressed X-ray-emitting gas surrounding the radio bubble is cool, and
does not provide evidence of recent shock heating by the radio lobe.

Fitting a number of different models to the total spectrum for the cluster
revealed that the best fit is achieved with a cooling flow model with a
mass deposition rate of $\dot{M} = 18.8^{+5.7}_{-4.5}~M_{\odot}$ 
yr$^{-1}$ and an elemental abundance of $0.66^{+0.064}_{-0.063}$ times the 
solar value.  No evidence is seen for excess absorption.  The mass deposition
rate is similar to that measured from previous observatories
({\it ASCA}, White 2000; {\it ROSAT}, Peres 1998), however we find that
the gas cools only over a limited temperature range, by a factor of three
from 2.7 to 0.9 keV.  This is similar to the result of Peterson et al.\ (2003)
who found $\dot{M} = 10\pm1~M_{\odot}$ in the range $kT$ to $kT/2$, 
where $kT = 2.1\pm0.2$ keV, using {\it XMM-Newton} data. 

The radio source in the center of Abell 262, with a 1.4 GHz power of
$P_{1.4} = 4.7 \times 10^{22}$ W Hz$^{-1}$, is orders of magnitude less 
luminous
than the central sources in other well-studied cooling flows such as
Hydra A (McNamara et al.\ 2000), Perseus (Fabian et al.\ 2000) and Abell 2052
(Blanton et al.\ 2001, 2003).  The cavity evacuated in the ICM by this
source, with a diameter of $\approx 5$ kpc, is much smaller than those 
measured in the other systems just mentioned, which have cavity diameters
of approximately 20 kpc.  The X-ray pressure measured in the bright shell
surrounding the cavity is higher than the 
radio equipartition pressure, as is found with the other sources.
Although the central gas pressures measured in Hydra A 
(McNamara et al.\ 2000), Perseus (Schmidt et al.\ 2002), and Abell
2052 (Blanton et al.\ 2001, 2003) are all similar to that in Abell 262,
the smaller bubble volume in A262 translates into much less energy being
injected into the cooling gas by the radio source than in the other systems.
A comparison of the energy injected by the radio source and the luminosity
of the cooling gas shows that the radio source energy is insufficient to
offset the cooling of the gas unless it is experiencing a less powerful
than average outburst, and was more powerful in the past.

\acknowledgements
We thank Joshua Kempner (CfA) and John Houck (MIT), who provided us with
their spectral mapping software which was used to create the temperature map.
Support for E. L. B. was provided by NASA through the {\it Chandra} Fellowship
Program, grant award number PF1-20017, under NASA contract number NAS8-39073.
This research was supported by the National Aeronautics and Space
Administration through $Chandra$ Award Numbers GO2-3160X and GO3-4155X,
issued by the $Chandra$ X-ray Observatory Center, which is operated by the
Smithsonian Astrophysical Observatory for and on behalf of NASA under
contract NAS8-39073.  Some support came from NASA {\it XMM-Newton} 
Grant NAG5-13089.


\vskip7.5truein
\includegraphics{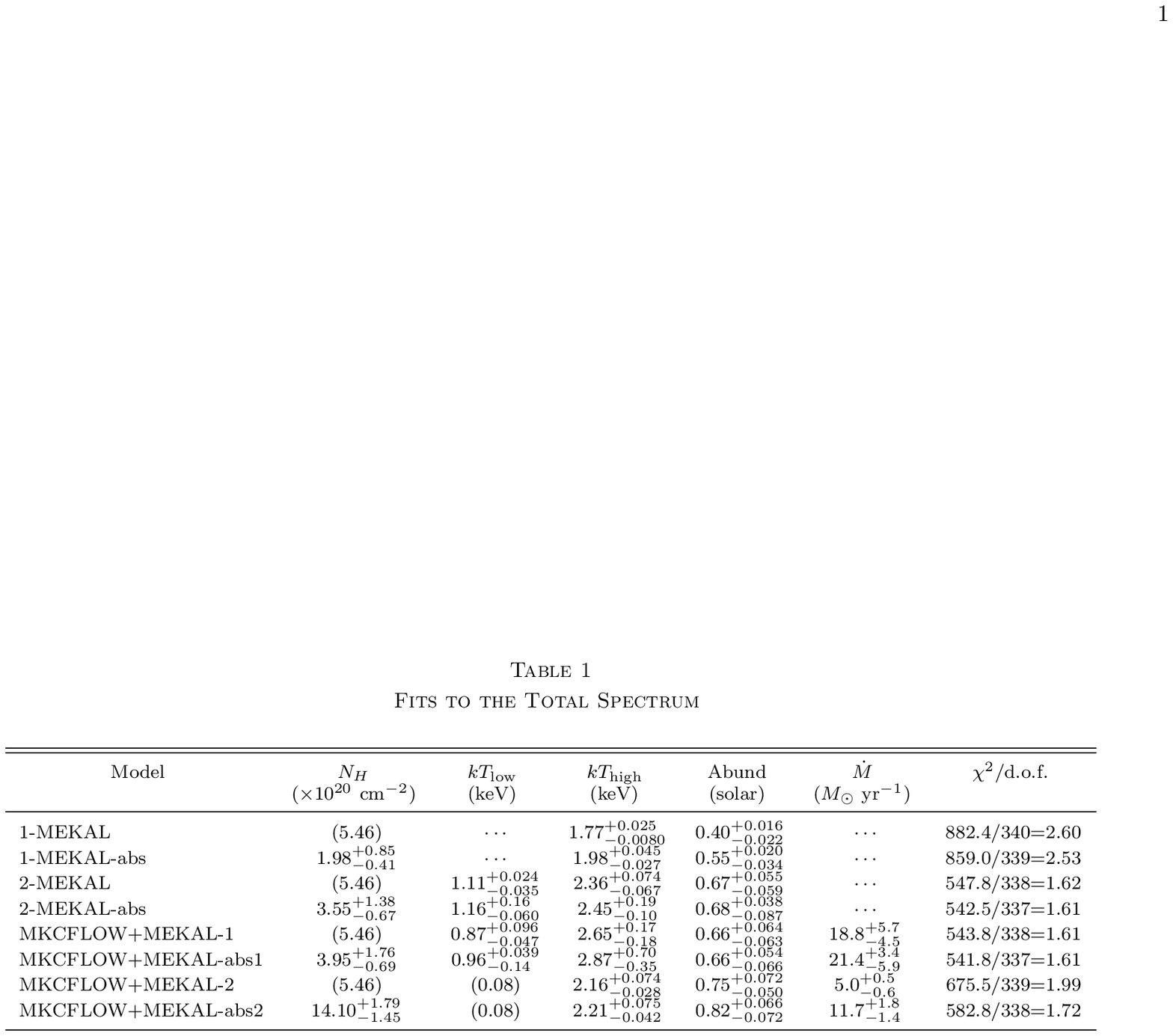}

\end{document}